\documentclass[final,numberedheadings,dvips]{aipproc}
\layoutstyle{8x11single}
\usepackage[latin1]{inputenc}
\usepackage[T1]{fontenc}
\usepackage{natbib}
\usepackage{graphicx}
\usepackage{color}
\usepackage{bm}
\usepackage{amsmath}
\usepackage{amssymb}
\usepackage{latexsym}

\newcommand{\Eqref}[1]{Eq.~(\ref{#1})}
\newcommand{\ii}{\mathrm{i}}
\newcommand{\nn}{\nonumber}
\newcommand{\be}{\begin{equation}}
\newcommand{\ba}{\begin{eqnarray}}
\newcommand{\ea}{\end{eqnarray}}
\newcommand{\ee}{\end{equation}}

\def \aspace{\!\!\!\!}
\def \openone{\hbox{$1\hskip -1.2pt\vrule depth 0pt height 1.6ex width 0.7pt\vrule depth 0pt height 0.3pt width 0.12em$}}

\begin{document}
\title{Nonlinear Terms of MHD Equations for Homogeneous Magnetized Shear Flow\thanks{This paper should be cited as published in the Proceedings of the School and Workshop on Space Plasma Physics (1--12 September 2010, Kiten, Bulgaria), AIP Conference Proceedings 1356, American Institute of Physics, Melville, NY, 2011.}}


\author{Z.~D.~Dimitrov}{
address={Department of Theoretical Physics, Faculty of Physics,
St.~Clement of Ohrid University at Sofia,\\
5 James Bourchier Blvd, BG-1164 Sofia, Bulgaria \\ e-mails: zlatan.dimitrov@gmail.com, maneva@linmpi.mpg.de, \\ tihomir.hristov@jhu.edu, tmishonov@phys.uni-sofia.bg}
}
\author{Y.~G.~Maneva}{}

\author{T.~S.~Hristov}{}

\author{T.~M.~Mishonov}{}

\begin{abstract}
We have derived the full set of MHD equations for incompressible shear flow of a magnetized fluid and
considered their solution in the wave-vector space. The linearized equations give the famous
amplification of slow magnetosonic waves and describe the magnetorotational instability.
The nonlinear terms in our analysis are responsible for the creation of turbulence and
self-sustained spectral density of the MHD (Alfv\'en and pseudo-Alfv\'en) waves.
Perspectives for numerical simulations of weak turbulence and calculation of the effective
viscosity of accretion disks are shortly discussed in k-space.
\end{abstract}
\maketitle

\section{Introduction}
%
How the chaotic matter did get organized into
compact astrophysical objects, such as stars,
when the Universe was created? How did the Sun's
rotation get slowed down (a central problem of
cosmogony)?  What is the physics behind quasars'
shining? The answers of all those unresolved
problems of contemporary physics go back to the
problem of the effective viscosity of weakly
magnetized plasmas in shear rotating flows.  For
half a century we have faced a kinetic
problem---how to calculate an effective viscosity---a
problem that is at the core of the machine for
making stars.  This longstanding problem has already
been approached in so many ways---any
proposal-writing astrophysicist has already
published his/her view and the literature of
analytical works and numerical simulations is
overwhelming. Yet the problem is still unsolved
and has not lost its attractiveness. Here we will
give it a try, too.
%

\section{Model and MHD Equations}
%
Our starting point are the conservation laws for energy and momentum for an incompressible fluid with mass density~$\rho$
\begin{eqnarray}
&&\partial_t(\rho V_i) + \partial_k(\Pi_{ik})=0\,, \\
&&\frac{\partial}{\partial t}\left(\frac{\rho V^2}{2} + \rho\tilde{\varepsilon} + \frac{B^2}{2\mu_0}\right)
+ \mathrm{div}\,\mathbf{q}=0\,,\\
&&\mathrm{div}\mathbf{V}=0, \qquad \rho=\mathrm{const},
\end{eqnarray}
where we have for total stress tensor $\Pi$ and heat flux $\mathbf{q}$ respectively
\begin{eqnarray}
\label{stress}
\Pi_{ik}\aspace&=&\aspace \rho V_iV_k + P\delta_{ik} - \eta\left(\frac{\partial V_i}{\partial x_k}+ \frac{\partial V_k}{\partial x_i}\right)
- \frac1{\mu_0}\left(B_iB_k -\frac12\delta_{ik}B^2\right),
\\
\label{heat}
\mathbf{q}\!\!\!\!\!&=&\!\!\!\!\!\rho\left(\frac{V^2}{2} +\tilde{w} \right)\mathbf{V} - \mathbf{V}\cdot{\sigma}'
-\kappa\nabla T
+\frac1{\mu_0}[\mathbf{B}\times(\mathbf{V}\times\mathbf{B})]
- \frac{\varepsilon_0c^2\varrho}{\mu_0}(\mathbf{B}\times\mathrm{rot}\,\mathbf{B}),
\end{eqnarray}
where $\tilde{\varepsilon}$ is the internal energy per unit mass, $\tilde{w}$ is the enthalpy per unit mass,
$\sigma_{ij}^{\prime}\equiv\eta(\partial_iV_k+\partial_kV_i)$ the viscous part of the stress tensor for an incompressible fluid,
$\eta$ is the viscosity, $\kappa$ is the heat conductivity, $T$ is the temperature, $\varrho$ is the Ohmic resistivity.
The formulas are written in SI, for a transition to Gaussian system we substitute $\mu_0=4\pi$ and $\varepsilon_0=1/4\pi,$
i.e., expressions are written in an invariant form.

For the magnetic field's energy density rate of change  we have
\begin{eqnarray}
\frac1{2\mu_0}\frac{\partial}{\partial t} \, B^2 =
\frac1{\mu_0}\mathbf{B}\cdot\frac{\partial \mathbf{B}}{\partial t} \aspace&=&\aspace
\frac1{\mu_0}\mathbf{B}\cdot[\nabla\times(\mathbf{V}\times\mathbf{B}) -\nu_\mathrm{m}\nabla\times(\nabla\times\mathbf{B})]\nonumber \\ \aspace&=&\aspace
\frac1{\mu_0}\nabla\cdot(\mathbf{B}\times(\mathbf{V}\times\mathbf{B}) -\nu_\mathrm{m}\mathbf{B}\times(\nabla\times\mathbf{B})).
\end{eqnarray}
We calculate the divergence of the total stress tensor \Eqref{stress}, and using that
\begin{eqnarray}
\partial_k\frac1{\mu_0}\left(B_iB_k -\frac12B^2\delta_{ik}\right)=\frac1{\mu_0}\left(\mathbf{B}\cdot\nabla\mathbf{B}
- \frac12\nabla B^2\right)=\frac1{\mu_0}(\mathbf{B}\times\mathrm{rot}\,\mathbf{B})=\mathbf{j}\times\mathbf{B},
\end{eqnarray}
we obtain the equation of motion for an incompressible plasma,
\begin{equation}
\rho\partial_t\mathbf{V}= -\mathbf{V}\cdot\nabla\mathbf{V} - \nabla P + \mathbf{j}\times\mathbf{B} + \eta\nabla^2\mathbf{V}.
\end{equation}
To close the set of equations we need to use Amp\`ere's law, $\nabla\times\mathbf{B}=\mu_0\mathbf{j},$
and generalized Ohm's law, $\mathbf{E}+\mathbf{V}\times\mathbf{B}=\eta\mathbf{j},$ and supplement them with Faraday's law
\begin{equation}
\frac{\partial \mathbf{B}}{\partial t}=-\nabla\times\mathbf{E}.
\end{equation}
For the second set of MHD equations we have
\begin{equation}
\frac{\partial \mathbf{B}}{\partial t}=\nabla\times\left(\mathbf{V}\times\mathbf{B}-\frac{\eta}{\mu_0}\nabla\times\mathbf{B} \right).
\end{equation}
The MHD equations for an incompressible fluid
$\rho=\mathrm{const},$ in homogeneous magnetic field $\mathbf{B}_0$,
shear flow with rate $A$, and angular velocity $\bm{\Omega}=A\omega\mathbf{e}_z=A\bm{\omega}$ are
\begin{eqnarray}
\label{MHD}
\rho \, \mathrm{D}_t \mathbf{V}
\aspace&=&\aspace - \nabla P + \left(\mathbf{j}
=\frac{\nabla \times \mathbf{B}}{\mu_0}\right)\times\mathbf{B}
-2\rho\,\bm{\Omega}\times\mathbf{V}
 + \rho\nu_\mathrm{k} \nabla^2\mathbf{V},\\
\label{MHD-B}
\mathrm{D}_t\mathbf{B}\aspace&=&\aspace\mathbf{B}\cdot\nabla\mathbf{V}+\nu_\mathrm{m}\nabla^2\mathbf{B},
\qquad \mathrm{div} \mathbf{V}=0, \qquad \mathrm{div} \mathbf{B}=0,
\end{eqnarray}
where $\mathrm{D}_t\equiv\partial_t+\mathbf V\cdot\nabla$ is the substantial
(convective) derivative,
$P$ is the pressure,  $\mathbf{j}$ is the current density and
$\nu_\mathrm{k}$ is the kinematic viscosity, The magnetic diffusivity
$\nu_\mathrm{m}=\varepsilon_0c^2\varrho$ is expressed by the Ohmic resistance $\varrho$ and
$\varepsilon_0=1/\mu_0c^2.$
In order to obtain a linear system of dimensionless MHD equations we use the following
\emph{Ansatz\/} for the velocity $\bf V$, the magnetic field $\bf B$, the wave vector $\bf Q$, and the pressure $P$
\begin{eqnarray}
\label{v_wave_Q}
 \mathbf{V}(t,\bf{r})\aspace&=&\aspace
\mathbf{V}_\mathrm{shear}(\mathbf{r})+\mathbf{V}_\mathrm{wave}(t,\mathbf{r}),
\qquad
\mathbf{V}_\mathrm{wave}=
\ii
V_\mathrm{A}\sum_{\mathbf{Q}}\mathbf{v}_{\mathbf{Q}}(\tau)\,\mathrm{e}^{\ii\mathbf{Q}\cdot\mathbf{X}},
\qquad
\mathbf{V}_\mathrm{shear}=Ax\mathbf{e}_y,
\\
\label{b_wave_Q}
\mathbf{B}(t,\bf{r})\aspace&=&\aspace\mathbf{B}_0+\mathbf{B}_\mathrm{wave}(t,\mathbf{r}),
\qquad
\mathbf{B}_\mathrm{wave}(t,\mathbf{r})=
B_0\sum_{\mathbf{Q}}\mathbf{b}_{\mathbf{Q}}(\tau)\,\mathrm{e}^{\ii\mathbf{Q}\cdot\mathbf{X}},\\
\label{p_wave_Q}
P(t,\mathbf{r})\aspace&=&\aspace P_0+ P_\mathrm{wave}(t,\mathbf{r}),
\qquad
P_\mathrm{wave}(t,\mathbf{r})=\rho V_\mathrm{A}^2\sum_{\mathbf{Q}}\mathcal{P}_{\mathbf{Q}}(\tau)\mathrm{e}^{\mathrm{i}\mathbf{Q}\cdot\mathbf{X}}.
\end{eqnarray}
where the sums are actually integrals with respect to 3-dimensional Eulerian wave-vector space with independent coordinates
$Q_x,\,Q_y,\,Q_z$
\begin{equation}
\sum_{\mathbf{Q}}=\int\int\int_{-\infty,\,-\infty,\,-\infty}^{+\infty,\,+\infty,\, +\infty}
\frac{\mathrm{d} Q_x\mathrm{d} Q_y \mathrm{d} Q_z}{(2\pi)^3}
=\int\mathrm{d}^3\left(\frac{\mathbf{Q}}{2\pi}\right)=\int\frac{\mathrm{d}^3{Q}}{(2\pi)^3},
\end{equation}
i.e., the sum is a short notation for Fourier integration with omitted differentials, integral limits and $2\pi$ multipliers.
For the static magnetic $\mathbf{B}_0$ field with magnitude
$B_0=\sqrt{B_{0y}^2+B_{0z}^2}$ we suppose a vertical $B_{0z}$ and an
azimuthal $B_{0y}$ components parameterized by an angle $\theta$ and
an unit vector $\bm{\alpha}$. We also assume that the Alfv\'en velocity
$V_{\mathrm{A}}$ is much smaller than the sound speed $c_{\mathrm{s}}$
\begin{equation}
\mathbf{B}_0=B_0\bm{\alpha},
\quad \bm{\alpha}=(0,\,\alpha_y=\sin\theta,\, \alpha_z=\cos\theta),
\quad \mathbf{V}_\mathrm{A}=\frac{\mathbf{B}_0}{\sqrt{\mu_0\rho}},
\quad V_\mathrm{A}=\frac{B_0}{\sqrt{\mu_0\rho}}.
\end{equation}
In the equations above we used the dimensionless space-vector
\be
\mathbf{X}=\left(\begin{array}{c} X\\Y\\Z \end{array} \right)
\equiv\frac{\mathbf{r}}{\Lambda}=\frac{A\mathbf{r}}{V_\mathrm{A}},\qquad
\Lambda\equiv\frac{V_\mathrm{A}}{A},\qquad \mathbf{r}=\left(\begin{array}{c} x\\y\\z \end{array} \right)
\ee
and its dimensionless wave-vector counterpart
$\mathbf{Q}$. Here, $\Lambda$ is the characteristic length of the system which we suppose
to be much smaller than space inhomogeneities, e.g., the accretion disk thickness.

\section{Wave-vector representation}
%
\subsection{Linear terms}
%
We have a space homogeneous physical system and indispensably its modes bear the character of plane waves.
The purpose of the present section is to find the Fourier transformation of the all the terms in the MHD equations (\ref{v_wave_Q}),
(\ref{b_wave_Q}) and (\ref{p_wave_Q}).

Let us start, for example, with the pressure. According to \Eqref{p_wave_Q} we have
\begin{eqnarray}
\frac{\nabla P}{\rho}\aspace&=&\aspace\frac{\ii V_\mathrm{A}^2}{\Lambda} \int
P_\mathbf{Q}(\tau)\mathbf{Q}\,\mathrm{e}^{\ii\mathbf{Q}\cdot\mathbf{X}}\frac{\mathrm{d}^3{Q}}{(2\pi)^3} =
\ii A V_\mathrm{A} \sum_\mathbf{Q} \mathbf{Q}P_\mathbf{Q}(\tau)\,\mathrm{e}^{\ii\mathbf{Q}\cdot\mathbf{X}}
\\
\mathcal{\hat{F}}(\frac{\nabla P}{\rho})\aspace&\equiv&\aspace \int \mathrm{e}^{-\ii\mathbf{Q}\cdot\mathbf{X}} \frac{\nabla P}{\rho}\,\mathrm{d}^3X
= \ii AV_\mathrm{A}\mathbf{Q}P_\mathbf{Q}(\tau).
\end{eqnarray}
Analogously, according to \Eqref{v_wave_Q} and \Eqref{b_wave_Q}, for the partial time derivatives we obtain
\begin{eqnarray}
&&\partial_t\mathbf{V}=
\ii A V_\mathrm{A}\sum_{\mathbf{Q}}\mathrm{e}^{\ii\mathbf{Q}\cdot\mathbf{X}}\partial_{\tau}\mathbf{v}_{\mathbf{Q}}(\tau), \qquad
\mathcal{\hat{F}} (\partial_t \mathbf{V})= iAV_\mathrm{A}\partial_\tau \mathbf{v}_\mathbf{Q}(\tau), \qquad
\label{time_V}\\
&&\partial_t\mathbf{B}= AB_0\sum_{\mathbf{Q}}\mathrm{e}^{\ii\mathbf{Q}\cdot\mathbf{X}}\partial_{\tau}\mathbf{b}_{\mathbf{Q}}(\tau),
\qquad
\mathcal{\hat{F}}(\partial_t\mathbf{B})=AB_0 \partial_\tau \mathbf{b}_\mathbf{Q}(\tau)\,.
\label{time_B}
\end{eqnarray}

More complicated are the Fourier transformations of the expressions, containing the shear flow $\mathbf{V}_\mathrm{shear}=V_\mathrm{A} X
\mathbf{e}_y$ and the wave variables
\begin{eqnarray}
\mathbf{V}_\mathrm{shear} \cdot \nabla \mathbf{V}_\mathrm{wave} \aspace&=&\aspace
-A V_\mathrm{A} \int \mathrm{e}^{\ii\mathbf{Q}\cdot\mathbf{X}}XQ_y \mathbf{v}(\tau) \frac{\mathrm{d}^3{Q}}{(2\pi)^3}\,,\\
\mathbf{V}_\mathrm{shear} \cdot \nabla \mathbf{B}_\mathrm{wave} \aspace&=&\aspace
\ii A B_0 \int \mathrm{e}^{\ii\mathbf{Q}\cdot\mathbf{X}} XQ_y \mathbf{b}(\tau) \frac{\mathrm{d}^3{Q}}{(2\pi)^3}\,.
\end{eqnarray}
Let variables $\mathbf{V}_\mathrm{wave}$ or $\mathbf{B}_\mathrm{wave}$ be presented by their Fourier components
$\psi(\mathbf{X})=\sum_\mathbf{Q}\mathrm{e}^{\mathrm{i} \mathbf{Q}\cdot \mathbf{X}}\psi_\mathbf{Q}$ and $\psi_\mathbf{Q}=
\mathcal{\hat{F}}(\psi(\mathbf{X}))$. Our task is to derive the Fourier transformation $\mathcal{\hat{F}}(\mathbf{X}\psi(\mathbf{X})).$
Using that $\mathbf{X} \mathrm{e}^{\mathrm{i} \mathbf{Q}\cdot \mathbf{X}}=-\mathrm{i}\partial_\mathbf{Q} \mathrm{e}^{\mathrm{i}
\mathbf{Q}\cdot \mathbf{X}}$  and the Gaussian theorem,
$\int_\mathcal{V}\mathrm{d}^3 Q \partial_\mathbf{Q}=\oint_{\partial\mathcal{V}}\mathrm{d}\mathbf{S}$
applied for the whole volume $\mathcal{V}$ in wave-vector space and its boundary $\partial\mathcal{V}$ we can make the partial
integration
\begin{equation}
 \mathbf{X}\psi(\mathbf{X})=\mathbf{X}\sum_\mathbf{Q}\mathrm{e}^{\mathrm{i} \mathbf{Q}\cdot \mathbf{X}}\psi_\mathbf{Q}
=-\mathrm{i}\sum_\mathbf{Q}\left\{
(\partial_\mathbf{Q} \mathrm{e}^{\mathrm{i} \mathbf{Q}\cdot \mathbf{X}})\psi_\mathbf{Q}=- \mathrm{e}^{\mathrm{i} \mathbf{Q}\cdot
\mathbf{X}}\partial_\mathbf{Q} \psi_\mathbf{Q}+\partial_\mathbf{Q}\left[\mathrm{e}^{\mathrm{i} \mathbf{Q}\cdot
\mathbf{X}}\partial_\mathbf{Q}\psi_\mathbf{Q}\right]\right\}=\sum_\mathbf{Q} \mathrm{e}^{\mathrm{i} \mathbf{Q}\cdot
\mathbf{X}}\ii\partial_\mathbf{Q}\psi_\mathbf{Q},
\end{equation}
in the limit $$\lim_{Q\rightarrow\infty}{(Q^3\psi_\mathbf{Q})}=0.$$
In such a way we derived the well-known in quantum mechanics operator representation $\hat{\mathbf{X}}=\ii\partial_\mathbf{Q}$
and derived the Fourier transformation
\begin{equation}
\mathcal{\hat{F}}(\mathbf{X}\psi(\mathbf{X}))=\ii \partial_\mathbf{Q}\psi_\mathbf{Q}.
\end{equation}
This expression is analogous to the Fourier transformation of the $\nabla$-operator
\begin{equation}
\mathcal{\hat{F}}(\nabla_{\!\!\mathbf{X}})=\ii\mathbf{Q},
\end{equation}
and yields
\be
\mathcal{\hat{F}}(X\mathbf{e}_y\cdot\nabla_{\!\!\mathbf{X}})=-Q_y\partial_{Q_x}.
\qquad \mathcal{\hat{F}}(\mathbf{V}_\mathrm{shear}\cdot\nabla)=-AQ_y\partial_{Q_x},
\qquad \mathbf{V}_\mathrm{shear} =V_\mathrm{A}X \mathbf{e}_y.
\label{baliga}
\ee
Those relations give that
\begin{equation}
\mathcal{\hat{F}}\left[\mathrm{D}_t^{\mathrm{\,shear}}\equiv\partial_t+\mathbf{V}_{\mathrm{shear}}\cdot\nabla=\partial_t+AX\partial_{_Y}
\right ]
=A\left\{\mathrm{D}_\tau^{\mathrm{\,shear}}\equiv\partial_\tau-Q_y\partial_{Q_x}
=\partial_\tau+\mathbf{U}_{\mathrm{shear}}(\mathbf{Q})\cdot\partial_{\mathbf{Q}}\right\}.
\end{equation}
In other words Fourier transformation of a linearized substantial derivative is again a linearized substantial derivative, but only in the wave-vector space.
For this purpose we introduced the field of shear flow in the wave-vector space
$\mathbf{U}_{\mathrm{shear}}(\mathbf{Q})\equiv-Q_y\mathbf{e}_x$; confer this result with
$\mathbf{V}_\mathrm{shear}/V_\mathrm{A} =X \mathbf{e}_y.$
Returning back to the velocity and magnetic field we arrive at
\ba
&&\mathcal{\hat{F}}[(\partial_t+\mathbf{V}_{\mathrm{shear}}\cdot\nabla)\mathbf{V}_{\mathrm{wave}}]
=\ii AV_\mathrm{A}[\partial_\tau+\mathbf{U}_{\mathrm{shear}}(\mathbf{Q})\cdot\partial_{\mathbf{Q}}]\mathbf{v}_\mathbf{Q}\,,\\
&&\mathcal{\hat{F}}[(\partial_t+\mathbf{V}_{\mathrm{shear}}\cdot\nabla)\mathbf{B}_{\mathrm{wave}}]
= AB_0[\partial_\tau+\mathbf{U}_{\mathrm{shear}}(\mathbf{Q})\cdot\partial_{\mathbf{Q}}]\mathbf{b}_\mathbf{Q}\,.
\ea
For the derivation of these equations we used \Eqref{baliga} and according to Eqs.~(\ref{time_V}) and (\ref{time_B})
$\mathcal{\hat{F}}(\partial_t)=A\partial_\tau.$

Very simple is the Fourier transformation of the dissipative terms which is reduced to the properties of the Laplacian
\begin{eqnarray}
&&\nu_\mathrm{k}\nabla^2\mathbf{V}=-\nu_\mathrm{k}\ii V_\mathrm{A}
\int\mathbf{v}_\mathbf{Q}(\tau)Q^2\mathrm{e}^{\ii\mathbf{Q}\cdot\mathbf{X}}\frac{\mathrm{d}^3{Q}}{(2\pi)^3}=
-\frac{\ii V_\mathrm{A}}{\Lambda} \nu_\mathrm{k} \int
\mathbf{v}_\mathbf{Q}(\tau)Q^2\mathrm{e}^{\ii\mathbf{Q}\cdot\mathbf{X}}\frac{\mathrm{d}^3{Q}}{(2\pi)^3}=
-\ii A V_\mathrm{A} \nu'_\mathrm{k} \int Q^2
\mathbf{v}_\mathbf{Q}(\tau)\mathrm{e}^{\ii\mathbf{Q}\cdot\mathbf{X}}\frac{\mathrm{d}^3{Q}}{(2\pi)^3}\,,\nonumber\\
&&\nu_m\nabla^2\mathbf{B}=-\frac{B_0}{\Lambda^2}\nu_m\int Q^2\mathbf{b}_\mathbf{Q}(\tau)\frac{\mathrm{d}^3{Q}}{(2\pi)^3}=
-AB_0\nu_m'\int Q^2\mathbf{b}_\mathbf{Q}(\tau)\frac{\mathrm{d}^3{Q}}{(2\pi)^3}\,,\nonumber\\
&&\frac1{\ii AV_\mathrm{A}}\mathcal{\hat{F}}(\nu\nabla^2\mathbf{V})=-\nu'_\mathrm{kin}Q^2\mathbf{v}(\tau),
\qquad\frac1{AB_0}\mathcal{\hat{F}}(\nu_\mathrm{m}\nabla^2\mathbf{B})= -\nu'_\mathrm{m}Q^2\mathbf{b}(\tau),
\qquad \nu_\mathrm{k}^\prime\equiv\frac{A}{V_\mathrm{A}^2}\nu_\mathrm{k}, \qquad
\nu_\mathrm{m}^\prime\equiv\frac{A}{V_\mathrm{A}^2}\nu_\mathrm{m}\,.
\end{eqnarray}
Hereafter for all terms coming from the velocity equation~(\ref{MHD}) we will separate a factor $\ii AV_\mathrm{A}$ and for all terms from
\Eqref{MHD-B} we will separate a factor $AB_0.$ Those factors will be common for the final equations in the wave-vector space.

For Coriolis force density per unit mass $-2\bm{\Omega}\times\mathbf{V}$, we derive
\begin{eqnarray}
&&-2\,\bm{\Omega}\times\mathbf{V} =
-2A\omega(-V_y\mathbf{e}_x+V_x\mathbf{e}_y)=
2A\omega\left(\begin{array}{c}
V_{y} \\
-V_{x} \\
0   \end{array}\right)=
2 A\omega \left(\begin{array}{c}
Ax + \ii V_\mathrm{A} \int v_{y,\mathbf{Q}}(\tau)\, \mathrm{e}^{\ii\mathbf{Q}\cdot\mathbf{X}}\frac{\mathrm{d}^3{Q}}{(2\pi)^3}\\
-\ii V_\mathrm{A} \int v_{x,\mathbf{Q}}(\tau)\, \mathrm{e}^{\ii\mathbf{Q}\cdot\mathbf{X}}\frac{\mathrm{d}^3{Q}}{(2\pi)^3}\\
0 \end{array}\right)\,,\nonumber\\
&&\mathcal{\hat{F}}(-2\bm{\Omega}\times\mathbf{V})=\ii AV_\mathrm{A}2\omega(v_{y,\mathbf{Q}}(\tau)\,\mathbf{e}_x -
v_{x,\mathbf{Q}}(\tau)\,\mathbf{e}_y).
\end{eqnarray}
The centrifugal term $2 \omega A^2 x$ is irrelevant for the wave amplitude equations.

Furthermore, we calculate Lorentz force per unit mass $\mathbf{j}\times\mathbf{B}$, and using $B_0^2/\mu_0=\rho V_\mathrm{A}^2$ we have
\begin{eqnarray}
\left(\left(\frac{\nabla\times\mathbf{B}}{\mu_0}\right)\times\frac{\mathbf{B}_0}{\rho}\right) \aspace&=&\aspace \frac{\ii B_0}
{\mu_0\rho\Lambda}\int(\mathbf{Q}\times\mathbf{b}_\mathbf{Q})\times(B_{0y}\mathbf{e}_y
+B_{0z}\mathbf{e}_z)\mathrm{e}^{\ii\mathbf{Q}\cdot\mathbf{X}} \frac{\mathrm{d}^3{Q}}{(2\pi)^3}
=\ii AV_\mathrm{A} \int\left[\left(\mathbf{Q}\times \mathbf{b}_\mathbf{Q}(\tau)\right) \times
\bm{\alpha}\right]\;\frac{\mathrm{d}^3{Q}}{(2\pi)^3}, \nonumber
\\
\mathcal{\hat{F}}\left(\left(\frac{\nabla\times\mathbf{B}}{\mu_0}\right)\times\frac{\mathbf{B}_0}{\rho}\right) \aspace&=&\aspace
\ii AV_\mathrm{A}\left[\left(\mathbf{Q}\times \mathbf{b}_\mathbf{Q}(\tau)\right) \times \bm{\alpha}\right]\,.
\end{eqnarray}

We have also other two zero terms having no influence on the wave dynamics. From momentum equation (\ref{MHD}) and the
magnetic field equation (\ref{MHD-B}) we have
\begin{eqnarray}
\mathbf{V}_\mathrm{shear} \cdot\nabla \mathbf{V}_\mathrm{shear} \aspace&=&\aspace Ax\,\mathbf{e}_y\cdot\nabla Ax\,\mathbf{e}_y
=A^2x\,\mathbf{e}_y\cdot \mathbf{e}_x \,\mathbf{e}_y=0\,,
\\
\mathbf{B}_0\cdot\nabla \mathbf{V}_\mathrm{shear} \aspace&=&\aspace AB_0(\alpha_y\mathbf{e}_y + \alpha_z\mathbf{e}_z)\cdot
\mathbf{e}_x\,\mathbf{e}_y=0\,.
\end{eqnarray}

For the last linear terms we have
\begin{eqnarray}
\mathbf{V}_\mathrm{wave}\cdot\nabla \mathbf{V}_\mathrm{shear}\aspace&=&\aspace
\ii AV_\mathrm{A}\int\mathbf{v_\mathbf{Q}(\tau)}\cdot
\mathbf{e}_{x}\,\mathbf{e}_{y}\,\mathrm{e}^{\ii\mathbf{Q}\cdot\mathbf{X}}\frac{\mathrm{d}^3Q}{(2\pi)^3},
\qquad
\mathcal{\hat{F}}(\mathbf{V}_\mathrm{wave}\cdot\nabla \mathbf{V}_\mathrm{shear})
=\ii AV_\mathrm{A} v_{x,\mathbf{Q}}(\tau)\mathbf{e}_y\,,
\\
\mathbf{B}_\mathrm{wave}\cdot\nabla \mathbf{V}_\mathrm{shear}\aspace&=&\aspace
AB_0 \int b_{x,\mathbf{Q}}(\tau)\,\mathbf{e}_y\,\mathrm{e}^{\ii
\mathbf{Q}\cdot\mathbf{X}}\frac{\mathrm{d}^3{Q}}{(2\pi)^3},
\qquad \quad\; \mathcal{\hat{F}}(\mathbf{B}_\mathrm{wave}\cdot\nabla \mathbf{V}_\mathrm{shear})= AB_0
b_{x,\mathbf{Q}}(\tau)\mathbf{e}_y\,,
\\
\mathbf{B}_0\cdot\nabla\mathbf{V}_\mathrm{wave}\aspace&=&\aspace\frac{\ii V_\mathrm{A} B_0}{\Lambda}\int (\bm{\alpha}
\cdot \ii \mathbf{Q})\mathbf{v}_\mathbf{Q}(\tau)\mathrm{e}^{\ii\mathbf{Q}\cdot\mathbf{X}}\frac{\mathrm{d}^3{Q}}{(2\pi)^3}, \quad
\mathcal{\hat{F}}(\mathbf{B}_0\cdot\nabla\mathbf{V}_\mathrm{wave})=
-AB_0 (\bm{\alpha}\cdot \mathbf{Q})\mathbf{v}_\mathbf{Q}(\tau)\,.
\end{eqnarray}
All linear terms are well-known from previous investigations of MHD waves in magnetized shear flows.
In the next subsection we will derive the nonlinear terms describing the wave--wave interaction coming from the convective time derivative.

\subsection{Nonlinear wave--wave interaction}
%
In order to derive the nonlinear term in the momentum equation (\ref{MHD}) we calculate
$\mathbf{V}_{\mathrm{wave}}\cdot\nabla\mathbf{V}_{\mathrm{wave}}$ using $\nabla
\mathrm{X}=\frac{A}{V_\mathrm{A}}\openone=\openone/\Lambda$
\begin{equation}
\mathbf{V}_{\mathrm{wave}}\cdot\nabla \mathbf{V}_{\mathrm{wave}}=
\sum_{\mathbf{Q}'}\ii V_\mathrm{A}\mathbf{v}(\tau,\mathbf{Q}')\mathrm{e}^{\ii\mathbf{Q}'\cdot\mathbf{X}}\cdot\nabla
\sum_{\mathbf{Q}''}\ii V_\mathrm{A}\mathbf{v}(\tau,\mathbf{Q}'')\mathrm{e}^{\ii\mathbf{Q}''\cdot\mathbf{X}}=
-\ii A V_\mathrm{A}\sum_{\mathbf{Q}'}\sum_{\mathbf{Q}''}\mathbf{v}_{\mathbf{Q}'}\cdot \mathbf{Q}'' \mathbf{v}_{\mathbf{Q}''}
\mathrm{e}^{\ii(\mathbf{Q}'+\mathbf{Q}'')\cdot\mathbf{X}}\,.
\end{equation}
For the sake of brevity in the last terms we will omit the time argument $\tau$ and write the wave-vector argument $\mathbf{Q}$ as index.
We make a Fourier transformation and obtain
\begin{eqnarray}&&
\hat{\mathcal{F}} \left(\mathbf{V}_{\mathrm{wave}}\cdot\nabla\mathbf{V}_{\mathrm{wave}}\right) \equiv
\int\mathrm{d}^3X
\mathrm{e}^{-\ii\mathbf{Q}\cdot\mathbf{X}}\left(\mathbf{V}_{\mathrm{wave}}\cdot\nabla\mathbf{V}_{\mathrm{wave}}\right)\\
\nonumber &&
=-\ii A V_\mathrm{A}\sum_{\mathbf{Q}'}\sum_{\mathbf{Q}''}\mathbf{v}_{\mathbf{Q}'}\cdot \mathbf{Q}''\mathbf{v}_{\mathbf{Q}''}
\delta\left(\frac{\mathbf{Q}'+\mathbf{Q}''-\mathbf{Q}}{2\pi}\right)
=-\ii A V_\mathrm{A}\sum_{\mathbf{Q}'}\mathbf{v}_{\mathbf{Q}-\mathbf{Q}'}\mathbf{v}_{\mathbf{Q}'}\cdot \mathbf{Q}.
\end{eqnarray}
The velocity and magnetic fields
\begin{equation}
\mathbf{V}_\mathrm{wave}(\tau,\mathbf{X})
 =\ii V_\mathrm{A}\int \mathrm{e}^{\mathbf{Q}\cdot\mathbf{X}}{\mathbf{v}}(\tau,\mathbf{Q})\frac{\mathrm{d}^3 X}{(2\pi)^3},
\qquad
\mathbf{B}_\mathrm{wave}(\tau,\mathbf{X})
 = B_0\int\mathrm{e}^{\mathbf{Q}\cdot\mathbf{X}}{\mathbf{b}}(\tau,\mathbf{Q})\frac{\mathrm{d}^3 X}{(2\pi)^3},
\end{equation}
have to be real, hence the Fourier components should be odd for the velocity and even for the magnetic field
\begin{equation}
 \mathbf{v}_{-\mathbf{Q}}=-\mathbf{v}_{\mathbf{Q}},
\qquad
 \mathbf{b}_{-\mathbf{Q}}=\mathbf{b}_{\mathbf{Q}}.
\end{equation}
Analogously, for the Fourier component of the wave-wave interaction part of the Lorentz force $\mathbf{j}_\mathrm{wave} \times
\mathbf{B}_\mathrm{wave}$ we obtain
\begin{equation}
\hat{\mathcal{F}}\left(\frac1{\mu_0\rho}(\nabla\times\mathbf{B}
_\mathrm{wave})\times\mathbf{B}_\mathrm{wave} \right)=
\int\mathrm{d}^3X\mathrm{e}^{-\ii\mathbf{Q}\cdot\mathbf{X}}\left(\frac1{\mu_0\rho}(\nabla\times\mathbf{B}
_\mathrm{wave})\times\mathbf{B}_\mathrm{wave} \right)
= \ii A V_\mathrm{A} \sum_\mathbf{Q'}\left(\mathbf{Q}'\times \mathbf{b}_{\mathbf{Q}'}\right)\times \mathbf{b}_{\mathbf{Q}-\mathbf{Q}'}.
\end{equation}

In such a way we derive the Fourier component of the nonlinear term of the momentum equation
\begin{eqnarray}
\mathbf{N}_{v,\mathbf{Q}}\aspace&\equiv&\aspace\frac1{\ii A
V_\mathrm{A}}\hat{\mathcal{F}}\left(-\mathbf{V}_{\mathrm{wave}}\cdot\nabla\mathbf{V}_{\mathrm{wave}}
+\frac1{\mu_0\rho}(\nabla\times\mathbf{B}_\mathrm{wave})\times\mathbf{B}_\mathrm{wave} \right)\\
\aspace&=&\aspace\sum_{\mathbf{Q}'}\left[\mathbf{v}_{\mathbf{Q}-\mathbf{Q}'}\mathbf{v}_{\mathbf{Q}'}\cdot\mathbf{Q}
+(\mathbf{Q}'\times\mathbf{b}_{\mathbf{Q}'})\times\mathbf{b}_{\mathbf{Q}-\mathbf{Q}'}\right].
\end{eqnarray}

Similarly, for the other nonlinear terms $\mathbf{B}_{\mathrm{wave}}\cdot\nabla \mathbf{V}_{\mathrm{wave}}$
and $\mathbf{V}_{\mathrm{wave}}\cdot\nabla\mathbf{B}_{\mathrm{wave}}$ we have
\begin{eqnarray}
\hat{\mathcal{F}}\left(\mathbf{B}_{\mathrm{wave}}\cdot\nabla\mathbf{V}_{\mathrm{wave}}\right)\aspace&=&\aspace
\int\mathrm{d}^3X
\mathrm{e}^{-\ii\mathbf{Q}\cdot\mathbf{X}}\left(\mathbf{B}_{\mathrm{wave}}\cdot\nabla\mathbf{V}_{\mathrm{wave}}\right)\\\nonumber
\aspace&=&\aspace-\frac{B_0V_\mathrm{A}}{\Lambda}\sum_{\mathbf{Q}'}\sum_{\mathbf{Q}''}\mathbf{b}_{\mathbf{Q}'}\cdot
\mathbf{Q}''\,\mathbf{v}_{\mathbf{Q}''}
\delta\left(\frac{\mathbf{Q}'+\mathbf{Q}''-\mathbf{Q}}{2\pi}\right)
=-AB_0\sum_{\mathbf{Q}'}\mathbf{b}_{\mathbf{Q}'}\cdot (\mathbf{Q}-\mathbf{Q}')\,\mathbf{v}_{\mathbf{Q}-\mathbf{Q}'},\\
\hat{\mathcal{F}}\left(\mathbf{V}_{\mathrm{wave}}\cdot\nabla\mathbf{B}_{\mathrm{wave}}\right)\aspace&=&\aspace
\int\mathrm{d}^3X
\mathrm{e}^{-\ii\mathbf{Q}\cdot\mathbf{X}}\left(\mathbf{V}_{\mathrm{wave}}\cdot\nabla\mathbf{B}_{\mathrm{wave}}\right)\\\nonumber
\aspace&=&\aspace-\frac{B_0V_\mathrm{A}}{\Lambda}\sum_{\mathbf{Q}'}\sum_{\mathbf{Q}''}\mathbf{v}_{\mathbf{Q}'}\cdot
\mathbf{Q}''\,\mathbf{b}_{\mathbf{Q}''}
\delta\left(\frac{\mathbf{Q}'+\mathbf{Q}''-\mathbf{Q}}{2\pi}\right)
=-AB_0\sum_{\mathbf{Q}'}\mathbf{v}_{\mathbf{Q}'}\cdot (\mathbf{Q}-\mathbf{Q}')\,\mathbf{b}_{\mathbf{Q}-\mathbf{Q}'}.
\end{eqnarray}
Those terms participate in the equation for the magnetic field. For their difference we have
\begin{eqnarray}
\mathbf{N}_{b,\mathbf{Q}}\aspace&\equiv&\aspace\frac1{AB_0}
\hat{\mathcal{F}}\left(\mathbf{B}_{\mathrm{wave}}\cdot\nabla\mathbf{V}_{\mathrm{wave}}-
\mathbf{V}_{\mathrm{wave}}\cdot\nabla\mathbf{B}_{\mathrm{wave}}\right)\\\nonumber
\aspace&=&\aspace\sum_{\mathbf{Q}'}\left[\mathbf{v}_{\mathbf{Q}'}\cdot (\mathbf{Q}-\mathbf{Q}')\,\mathbf{b}_{\mathbf{Q}-\mathbf{Q}'}
-\mathbf{b}_{\mathbf{Q}'}\cdot (\mathbf{Q}-\mathbf{Q}')\,\mathbf{v}_{\mathbf{Q}-\mathbf{Q}'}\right]\\\nonumber
\aspace&=&\aspace-\mathbf{Q}\times\sum_{\mathbf{Q}'}(\mathbf{v}_{\mathbf{Q}'} \times \mathbf{b}_{\mathbf{Q}-\mathbf{Q}'})
\end{eqnarray}
As the function in $r$-space
\begin{equation}
\mathrm{rot}\left(\mathbf{V}_\mathrm{wave}\times\mathbf{B}_{\mathrm{wave}}\right)=
\mathbf{B}_{\mathrm{wave}}\cdot\nabla\mathbf{V}_{\mathrm{wave}}-
\mathbf{V}_{\mathrm{wave}}\cdot\nabla\mathbf{B}_{\mathrm{wave}}
\end{equation}
has zero divergence
\begin{equation}
\mathrm{div}\left[\mathrm{rot}\left(\mathbf{V}_\mathrm{wave}\times\mathbf{B}_{\mathrm{wave}}\right)\right]=0
\end{equation}
its Fourier transform is transversal $\mathbf{Q}\cdot\mathbf{N}_{b,\mathbf{Q}}=0$ and automatically
$\mathbf{N}_{b,\mathbf{Q}}^\perp=\mathbf{N}_{b,\mathbf{Q}}.$

In order to merge the so derived nonlinear terms in a next subsection we will rederive the
linear terms in the Lagrangian wave-vector space.

\section{Elimination of pressure in the final MHD equations}
%
It is common in  MHD to formally seek the limit of a particular expression
for infinite sound speed $c_\mathrm{s}\rightarrow \infty.$
Due to the complexity of the problem this standard approach for consideration weak magnetic fields when $V_\mathrm{A}\ll c_\mathrm{s}$
is inapplicable to our problem and we have to look for direct elimination of the pressure.
After substituting Fourier transformations in Eqs.~(\ref{MHD}) and (\ref{MHD-B}) we obtain
\begin{eqnarray}
\left(\partial_\tau + \mathbf{U}_\mathrm{shear}\cdot\partial_\mathbf{Q} \right)\mathbf{v}_\mathbf{Q} \aspace&=&\aspace
-v_{x,\mathbf{Q}}\mathbf{e}_y +\mathbf{Q}P_\mathbf{Q} +\left[(\mathbf{Q}\times\mathbf{b}_\mathbf{Q})\times\bm{\alpha}\right]
+ 2\omega(v_{y,\mathbf{Q}}\mathbf{e}_x - v_{x,\mathbf{Q}}\mathbf{e}_y)- \nu'_\mathrm{k}Q^2\mathbf{v}_\mathbf{Q}
\\
&\quad&\,+\sum_{\mathbf{Q}'}\left[\mathbf{v}_{\mathbf{Q}-\mathbf{Q}'}\mathbf{v}_{\mathbf{Q}'}\cdot\mathbf{Q}
+(\mathbf{Q}'\times\mathbf{b}_{\mathbf{Q}'})\times\mathbf{b}_{\mathbf{Q}-\mathbf{Q}'}\right] \nonumber,
\\
\left(\partial_\tau + \mathbf{U}_\mathrm{shear}\cdot\partial_\mathbf{Q} \right)\mathbf{b}_\mathbf{Q} \aspace&=&\aspace b_{x,\mathbf{Q}}\mathbf{e}_y
-(\mathbf{Q}\cdot\bm{\alpha})\mathbf{v}_\mathbf{Q} -\nu'_\mathrm{m}Q^2\mathbf{b}_\mathbf{Q}
-\mathbf{Q}\times\sum_{\mathbf{Q}'}(\mathbf{v}_{\mathbf{Q}'} \times \mathbf{b}_{\mathbf{Q}-\mathbf{Q}'}).
\end{eqnarray}
For the sake of brevity we introduce
\begin{eqnarray}
\label{force_short}
\mathcal{F}_\mathbf{Q}\equiv Q_y\frac{\partial \mathbf{v}_\mathbf{Q}}{\partial Q_x}-\mathbf{e}_y \mathbf{e}_x \cdot \mathbf{v}_{\mathbf{Q}}
+ [(\mathbf{Q}\times\mathbf{b})\times\bm{\alpha}] + 2\bm{\omega}\times\mathbf{v}_\mathbf{Q} + \nu'_\mathrm{k}Q^2\mathbf{v}_\mathbf{Q}
+\sum_{\mathbf{Q}'}\left[\mathbf{v}_{\mathbf{Q}-\mathbf{Q}'}\mathbf{v}_{\mathbf{Q}'}\cdot\mathbf{Q}
+(\mathbf{Q}'\times\mathbf{b}_{\mathbf{Q}'})\times\mathbf{b}_{\mathbf{Q}-\mathbf{Q}'}\right].
\end{eqnarray}
Then the equation for the velocity can be rewritten as
\begin{equation}
\label{short_v}
\partial_\tau \mathbf{v}_\mathbf{Q}=P_\mathbf{Q}\mathbf{Q} +\mathcal{F}_\mathbf{Q}.
\end{equation}
In order to express the pressure, we multiply both sides of this equation by $\mathbf{Q}$
\be
\partial_\tau (\mathbf{Q}\cdot\mathbf{v}_\mathbf{Q})= Q^2P_\mathbf{Q}+\mathbf{Q}\cdot\mathcal{F}_\mathbf{Q}.
\ee
The incompressibility condition $\mathbf{Q}\cdot\mathbf{v}_\mathbf{Q}=0$ gives for the pressure the solution of the Poisson equation
\begin{eqnarray}
\mathcal{P}=-\frac{\mathbf{Q} \cdot \mathcal{F}_\mathbf{Q}}{Q^2}
\aspace&=&\aspace-\frac{1}{Q^2}\left\{ 2\mathbf{Q}\cdot\mathbf{e}_y \mathbf{e}_x \cdot\mathbf{v}_\mathbf{Q}
+2\bm{\omega}\times\mathbf{v}_\mathbf{Q}
+\mathbf{Q}\cdot\left[\left(\mathbf{Q}\times\mathbf{b}_\mathbf{Q}\right)\times\bm{\alpha}\right]
\right\} \nonumber\\
&&-\frac{1}{Q^2}
\sum_{\mathbf{Q}'}\left\{\mathbf{Q}\cdot\mathbf{v}_{\mathbf{Q}-\mathbf{Q}'}\mathbf{v}_{\mathbf{Q}'}\cdot\mathbf{Q}
+\left[(\mathbf{Q}'\times\mathbf{b}_{\mathbf{Q}'})\times\mathbf{b}_{\mathbf{Q}-\mathbf{Q}'}\right]\cdot\mathbf{Q}\right\},
\end{eqnarray}
where we used the obvious vector relations
\be
\mathbf{Q}\cdot\mathbf{e}_y \mathbf{e}_x \cdot\mathbf{v}_\mathbf{Q}= 2v_xQ_y,
\qquad \mathbf{Q}\cdot(\bm{\omega}\times\mathbf{v}_\mathbf{Q})=\omega(Q_yv_x-Q_xv_y),
\qquad \left(\mathbf{Q}\times\mathbf{b}_\mathbf{Q}\right)\times\bm{\alpha}
=(\mathbf{Q}\cdot\bm{\alpha})(\mathbf{Q}\cdot\mathbf{b}) - \mathbf{Q}^2(\mathbf{b}\cdot\bm{\alpha})\,.
\ee

This formula for the pressure we substitute in \Eqref{short_v} which takes the form
\begin{equation}
\label{uskorenie}
\partial_\tau \mathbf{v}_\mathbf{Q}=\mathcal{F}_\mathbf{Q}^{\perp}
=\mathcal{F}_\mathbf{Q} -\frac{\mathbf{Q}\otimes \mathbf{Q}}{Q^2}\mathcal{F}_\mathbf{Q}=\Pi^{\perp\mathbf{Q}}\mathcal{F}_\mathbf{Q},
\end{equation}
where
\be
\Pi^{\perp\mathbf{Q}}\equiv\openone-\frac{\mathbf{Q}\otimes \mathbf{Q}}{Q^2}=\openone-\mathbf{n}\otimes \mathbf{n},
\qquad \mathbf{n}\equiv \frac{\mathbf{Q}}{Q}
\ee
is the projection operator which applies to the part of a vector, perpendicular to the wave-vector.
In other words, the elimination of the pressure conserves the perpendicular part of
the Fourier component of the force $\mathcal{F}_\mathbf{Q}$ in the used dimensionless variables.
Equation (\ref{uskorenie}) means that the velocity field remains orthogonal to the wave vector.
If in the beginning $\mathbf{Q}\cdot\mathbf{v}_\mathbf{Q}(\tau_0)=0$, the evolution gives that
$\mathbf{Q}\cdot\mathbf{v}_\mathbf{Q}(\tau)=0$ for every $\tau>\tau_0.$

Using that for the velocity as applicable for every orthogonal vector
\be
\Pi^{\perp}\partial_\tau\mathbf{v}_\mathbf{Q}=\partial_\tau\mathbf{v}_\mathbf{Q},
\qquad \Pi^{\perp}\mathbf{v}_\mathbf{Q}=\mathbf{v}_\mathbf{Q}
\ee
we can rewrite \Eqref{uskorenie} as
\be
\Pi^{\perp}(\partial_\tau\mathbf{v}_\mathbf{Q} - \mathcal{F}_\mathbf{Q})=0.
\ee

In order to take into account the $Q_y\partial \mathbf{v}_\mathbf{Q}/\partial Q_x$ term in \Eqref{force_short}
we use the obvious relations
\begin{equation}
Q_y\frac{\partial}{\partial Q_x}\left(\mathbf{v}_\mathbf{Q} \cdot \frac{\mathbf{Q}\mathbf{Q}}{Q^2}\right)=
Q_y\frac{\partial \mathbf{v}_\mathbf{Q}}{\partial Q_x} \cdot \frac{\mathbf{Q}\mathbf{Q}}{Q^2}+
\frac{Q_y v_{x,\mathbf{Q}}}{Q^2} \mathbf{Q}=0,
\qquad \mathbf{v}_\mathbf{Q} \cdot \mathbf{Q}=0.
\end{equation}
Now we represent the projection of the advective term $\mathbf{U}_\mathrm{shear}\cdot\partial_\mathbf{Q}\mathbf{v}_\mathbf{Q}$ as
\begin{eqnarray}
\Pi^{\perp\mathbf{Q}}(\mathbf{U}_\mathrm{shear}\cdot\partial_\mathbf{Q}\mathbf{v}_\mathbf{Q})
=-Q_y\frac{\partial \mathbf{v}_\mathbf{Q}}{\partial Q_x}-n_y\mathbf{n}v_{x,\mathbf{Q}}
\end{eqnarray}
and arrive at the momentum equation in the form where the projection operator exists explicitly only in the nonlinear term
\begin{eqnarray}
\left(\partial_\tau + \mathbf{U}_\mathrm{shear}\cdot\partial_\mathbf{Q} \right)\mathbf{v}_\mathbf{Q}(\tau)\aspace &=&\aspace
-v_{x,\mathbf{Q}}\mathbf{e}_y+2n_y\mathbf{n}v_{x,\mathbf{Q}}
+ 2\omega\mathbf{n}(n_yv_{x,\mathbf{Q}}-n_xv_{y,\mathbf{Q}})
+2\bm{\omega}\times v_{\mathbf{Q}} + (\bm{\alpha}\cdot\mathbf{Q})\mathbf{b}_\mathbf{Q} \nonumber\\
\aspace&&\aspace-\nu^\prime_kQ^2\mathbf{v}_\mathbf{Q}+\Pi^{\perp\mathbf{Q}}\sum_{\mathbf{Q}'}\left[\mathbf{v}_{\mathbf{Q}
-\mathbf{Q}'}\mathbf{v}_{\mathbf{Q}'}\cdot\mathbf{Q}
+(\mathbf{Q}'\times\mathbf{b}_{\mathbf{Q}'})\times\mathbf{b}_{\mathbf{Q}-\mathbf{Q}'}\right],
\\
\label{MagneticEvolution}
\left(\partial_\tau + \mathbf{U}_\mathrm{shear}\cdot\partial_\mathbf{Q} \right)\mathbf{b}_\mathbf{Q}(\tau) \aspace&=&\aspace b_{x,\mathbf{Q}}\mathbf{e}_y
-(\mathbf{Q}\cdot\bm{\alpha})\mathbf{v}_\mathbf{Q} -\nu'_\mathrm{m}Q^2\mathbf{b}_\mathbf{Q}
-\mathbf{Q}\times\sum_{\mathbf{Q}'}(\mathbf{v}_{\mathbf{Q}'} \times \mathbf{b}_{\mathbf{Q}-\mathbf{Q}'}),\\
&&\mathbf{v}_\mathbf{Q}(\tau_0)=\Pi^{\perp}\mathbf{v}_\mathbf{Q}(\tau_0),
\qquad
\mathbf{b}_\mathbf{Q}(\tau_0)=\Pi^{\perp}\mathbf{b}_\mathbf{Q}(\tau_0).
\end{eqnarray}
For numerical calculations the incompressibility conditions
$\mathbf{n}\cdot \mathbf{b}_\mathbf{Q}=0$ and $\mathbf{n}\cdot \mathbf{v}_\mathbf{Q}=0$ can be used as a criterion for the error.

Using the relation
\be
[\mathbf{U}_\mathrm{shear}\cdot\partial_\mathbf{Q}\mathbf{b}_\mathbf{Q}(\tau)]\cdot\mathbf{Q}= Q_yb_{x,\mathbf{Q}},
\ee
one can easily check that the equation for the evolution of the magnetic field, \Eqref{MagneticEvolution}, can
also be presented as the evolution of its part, perpendicular to the wave-vector
\begin{equation}
\partial_\tau\mathbf{b}_\mathbf{Q}  + \Pi^{\perp}\mathbf{U}_\mathrm{shear}\cdot\partial_\mathbf{Q} \mathbf{b}_\mathbf{Q}
= \Pi^{\perp} \mathbf{e}_y\mathbf{e}_x\cdot\mathbf{b}_{\mathbf{Q}} -(\mathbf{Q}\cdot\bm{\alpha})\mathbf{v}_\mathbf{Q}
-\nu'_\mathrm{m}Q^2\mathbf{b}_\mathbf{Q}
-\mathbf{Q}\times\sum_{\mathbf{Q}'}(\mathbf{v}_{\mathbf{Q}'} \times \mathbf{b}_{\mathbf{Q}-\mathbf{Q}'}).
\end{equation}
Together with $\mathbf{v}_\mathbf{Q}=\Pi^{\perp}\mathbf{v}_\mathbf{Q}$ this equation automatically
gives $\mathbf{b}_\mathbf{Q}=\Pi^{\perp}\mathbf{b}_\mathbf{Q}$ and $\mathrm{div} \mathbf{B}=0.$

In the matrix form the set of MHD equations reads as
\begin{equation}
\mathrm{D}_\tau\Psi=\mathsf{M}\Psi + \mathsf{N},
\end{equation}
where
\begin{equation}
\mathsf{N}=\left(\begin{array}{c}
\mathbf{N}_{b,\mathbf{Q}}^{\perp}\\
\mathbf{N}_{v,\mathbf{Q}}^{\perp}
\end{array}\right)
=
\left(\begin{array}{c}
-\mathbf{Q}\times\sum_{\mathbf{Q}'}(\mathbf{v}_{\mathbf{Q}'} \times \mathbf{b}_{\mathbf{Q}-\mathbf{Q}'})\\
\Pi^{\perp\mathbf{Q}}\sum_{\mathbf{Q}'}\left[\mathbf{v}_{\mathbf{Q}-\mathbf{Q}'}\mathbf{v}_{\mathbf{Q}'}\cdot\mathbf{Q}
+(\mathbf{Q}'\times\mathbf{b}_{\mathbf{Q}'})\times\mathbf{b}_{\mathbf{Q}-\mathbf{Q}'}\right]
                 \end{array}\right)\,,
\end{equation}
and
\begin{small}
\begin{equation}
\mathsf{M} =
\left(\begin{array}{ccc|ccc}
-\nu^\prime_\mathrm{m}Q^2 & 0 & 0 & -Q_\alpha& 0 & 0\\
1 & -\nu^\prime_\mathrm{m}Q^2 & 0 & 0 & -Q_\alpha & 0 \\
0 & 0& -\nu^\prime_\mathrm{m}Q^2  & 0 & 0 & -Q_\alpha\\
\hline
Q_\alpha& 0 & 0 & 2n_yn_x(\omega+1)-\nu^\prime_\mathrm{k}Q^2 & -2n_xn_x\omega+2\omega & 0 \\
0 & Q_\alpha & 0 &2n_yn_y(\omega+1) -(2\omega+1) & -2n_xn_y\omega -\nu^\prime_\mathrm{k}Q^2 & 0 \\
0 & 0 & Q_\alpha & 2n_yn_z(\omega+1) & -2n_xn_z \omega  & -\nu^\prime_\mathrm{k}Q^2
\end{array}\right),\qquad
\Psi_\mathbf{Q}=\left(\begin{array}{c}
            b_x\\
	    b_y\\
	    b_z\\
	    v_x\\
	    v_y\\
	    v_z
           \end{array}\right)
\end{equation}
\end{small}
with $Q_\alpha\equiv\mathbf{Q}\cdot\bm{\alpha}.$

The matrix can also be represented as
\begin{equation}
\mathsf{M} =
\left(\begin{array}{c|c}
\mathsf{M}_{bb} & \mathsf{M}_{bv}\\
\hline
\mathsf{M}_{vb} & \mathsf{M}_{vv} \\
\end{array}\right),\qquad
\Psi_\mathbf{Q}=\left(\begin{array}{c}
            \mathbf{b}\\
	    \mathbf{v}
	    \end{array}\right),
\end{equation}
\begin{equation}
\mathsf{M}_{vv} =
2n_y \left(\begin{array}{ccc}
n_x & 0 & 0\\
n_y & 0 & 0\\
n_z & 0 & 0
\end{array}\right)
-
\left(\begin{array}{ccc}
0 & 0 & 0\\
1 & 0 & 0\\
0 & 0 & 0
\end{array}\right)
+2\omega
\left(\begin{array}{ccc}
n_xn_y & (n_y^2+n_z^2) & 0\\
-(n_x^2+n_z^2) & -n_xn_y & 0\\
n_yn_z & -n_xn_z & 0
\end{array}\right)
-\nu^\prime_\mathrm{k}Q^2\openone,
\end{equation}
\begin{equation}
\mathsf{M}_{vb} = Q_\alpha\openone, \qquad \mathsf{M}_{bv} = -Q_\alpha\openone, \qquad
\mathsf{M}_{bb} =
\left(\begin{array}{ccc}
0 & 0 & 0\\
1 & 0 & 0\\
0 & 0 & 0
\end{array}\right)
-\nu^\prime_\mathrm{m}Q^2\openone.
\end{equation}
With the help of the matrices in this representation, in the next section we will make Lyapunov analysis
of the linearized MHD equations.

\section{Lyapunov analysis of the linearized set of equations in Lagrangian variables}
\label{sec:Lyapunov}
%
For small $Q_y$ we may neglect the advective term $\mathbf{U}_\mathrm{shear}\cdot\partial_\mathbf{Q}=-Q_y\partial_{Q_x}.$
Then the linearized MHD equations take the form
\be
\mathrm{D}_\tau\Psi=\mathsf{M}\Psi, \qquad (\mathbf{Q},\mathbf{Q})\cdot\Psi=0.
\ee
To perform an instability analysis we make use of the exponential substitution $\Psi=\exp(\lambda\tau)\psi$
which leads to an eigenvalue problem with transversality conditions
\be
\mathsf{M}(\mathbf{Q})\Psi_\mathbf{Q}=\lambda\Psi_\mathbf{Q},
\qquad \mathbf{Q}\cdot\mathbf{b}_\mathbf{Q}=0=\mathbf{Q}\cdot\mathbf{v}_\mathbf{Q};
\ee
to make it short we can further omit the index $\mathbf{Q}$
\be
\left(\mathsf{M}-\lambda\openone\right)\Psi=0,
\qquad \mathbf{Q}\cdot\mathbf{b}=0=\mathbf{Q}\cdot\mathbf{v}.
\ee
Should we substitute the incompressibility and transversality conditions
\be
\label{transversal}
v_z=-\frac{Q_x v_x+Q_y v_y}{Q_z}, \qquad b_z=-\frac{Q_x b_x+Q_y b_y}{Q_z},
\ee
in the secular equation, we would end up with an overdetermined set of equations.
To avoid it, we omit the equations which initially have $\lambda b_z$ and $\lambda v_z$ terms, i.e., the 3-rd and the 6-th rows in the secular equation.
In such a way we derive a secular equation for a reduced matrix
\begin{small}
\begin{equation}
\left(\tilde{\mathsf{M}}-\lambda\openone\right)\psi=0,
\quad
\psi=\left(\begin{array}{c}
            b_x\\
	        b_y\\
	        v_x\\
	        v_y
                      \end{array}
                \right),
\quad \mathsf{\tilde{M}} =
\left(\begin{array}{cc|cc}
-\nu^\prime_\mathrm{m}Q^2 & 0 & -Q_\alpha& 0 \\
1 & -\nu^\prime_\mathrm{m}Q^2 & 0 & -Q_\alpha  \\
\hline
Q_\alpha& 0 & 2n_yn_x(\omega+1)-\nu^\prime_\mathrm{k}Q^2 & -2n_xn_x\omega+2\omega  \\
0 & Q_\alpha &2n_yn_y(\omega+1) -(2\omega+1) & -2n_xn_y\omega -\nu^\prime_\mathrm{k}Q^2
\end{array}\right).
\end{equation}
\end{small}
$\!\!$This secular equation
\be
P_4(\lambda; \mathbf{Q},\nu_\mathrm{m}^{\,\prime},\nu_\mathrm{k}^{\,\prime})
\equiv\mathrm{det}\left(\tilde{\mathsf{M}}-\lambda\openone\right)=0
\ee
has 4 eigenvalues and via a calculation of the eigenvectors, we can derive $b_z$ and $v_z$ according to the transversality conditions given by \Eqref{transversal}.

For an ideal fluid both $\nu_\mathrm{m}^{\,\prime}$ and $\nu_\mathrm{k}^{\,\prime}$ are equal to zero. Omitting the viscosity terms we have a relatively simple form for the secular equation
\ba
\nn
P_4(\lambda; \mathbf{Q},\nu_\mathrm{m}^{\,\prime}=0,\nu_\mathrm{k}^{\,\prime}=0)
&\!\!\!=\!\!\!&
\lambda^4
-2n_yn_x\lambda^3
+\left\{\left[(4-8n_y^2)n_x^2+4-4n_y^2\right]\omega_{_\mathrm{C}}^2 +
\left[(2-8n_y^2)n_x^2-4n_y^2+2\right]\omega_{_\mathrm{C}}+2Q_\alpha^2\right\}\lambda^2\\
&&{}-2Q_\alpha^2n_yn_x\lambda
+ 2Q_\alpha^2(n_x^2+1)\omega_{_\mathrm{C}}+Q_\alpha^4=0.
\ea
As we pointed out these eigenvalues give only a WKB approximation for the dynamics of MHD variables $\psi(\tau).$
For the special case of $Q_y=0$, which corresponds to an axial-symmetric motion,
with a rotation along the $z$-axis, the secular equation gives directly the growth rates of the linearized MHD equations.
\be
P_4(\lambda; Q_y=0, \nu_\mathrm{m}^{\,\prime}=0,\nu_\mathrm{k}^{\,\prime}=0)=
\lambda^4+
2\left[Q_\alpha^2 + (1 +2\omega_{_\mathrm{C}})(n_x^2+1)\omega_{_\mathrm{C}}\right]\lambda^2
+2Q_\alpha^2(n_x^2+1)\omega_{_\mathrm{C}}+Q_\alpha^4=0 .
\ee
The most restricted case is for the wave-vectors parallel to the rotation axis
$\mathbf{Q}=Q\mathbf{e}_z$ when $Q_\alpha=Q_z \cos{\theta}$
\be
\label{MRI}
P_4(\lambda; Q_x=0, Q_y=0, \nu_\mathrm{m}^{\,\prime}=0,\nu_\mathrm{k}^{\,\prime}=0)=
\lambda^4 +2\left[Q_\alpha^2 + (1 +2\omega_{_\mathrm{C}})\omega_{_\mathrm{C}}\right]\!\lambda^2
+ \left(Q_\alpha^2+2\omega_{_\mathrm{C}}\right)\!Q_\alpha^2=0.
\ee
This is perhaps the most cited bi-quadratic equation in the whole history of science because it describes the magnetorotational instability (MRI) discovered by Velikhov~\cite{Velikhov:59} in 1959.
In the astrophysics, this equation was recognized and overexposed by many astrophysical grants 30 years later, see equation Eq.~(111) of Ref.~\cite{Balbus:98} and historical remarks therein.
If we consider the special case of pure shear $\omega_{_\mathrm{C}}=0$ with $Q_y=0$
this dispersion equation gives the usual Alfv\'en waves
\be
(\lambda^2 + Q_\alpha^2)^2 =0,\qquad \omega=\left|Q_\alpha\right|,
\ee
i.e., the rotation destabilizes the Alfv\'en waves. The polarization of the magnetic field and the velocity of the Alfv\'en waves are along the shear flow.

For pure axial magnetic field $\mathbf{B}=B\mathbf{e}_z,$ i.e., $\bm{\alpha}=(0,\,0,\,1),$ and $Q_\alpha=Q_z.$ The matrix reduction is then given by simply erasing the $z$-components and taking into account only the $x$- and $y$-projections of the equations of motions
\begin{equation}
\tilde{\mathsf{M}}_\mathrm{MRI} =
\left(\begin{array}{cc|cc}
 0 & 0 & -Q_z& 0 \\
 1 & 0 & 0 & -Q_z \\
\hline
Q_z & 0 & 0 & 2\omega_{_\mathrm{C}}  \\
0 & Q_z & -(2\omega_{_\mathrm{C}}+1) & 0 \\
\end{array}\right),\qquad
\psi=\left(\begin{array}{c}
            b_x\\
	    b_y\\
	    v_x\\
	    v_y
           \end{array}\right).
\end{equation}
The secular equation is equation (\ref{MRI}) for the MRI with $Q_\alpha=Q_z.$

The projection method can be generalized in the general case if
we introduce 2 unit vectors perpendicular to the wave-vector $\mathbf{e}_\mathbf{Q}=\mathbf{Q}/Q$
\ba
&\left|2\right>\aspace&\aspace=\mathbf{e}_2=\frac{\mathbf{e}_z\times\mathbf{e}_\mathbf{Q}}{|\mathbf{e}_z\times\mathbf{e}_\mathbf{Q}|}=
\frac1{\sqrt{Q_x^2+Q_y^2}}\left(\begin{array}{c} -Q_y\\Q_x \\0  \end{array}\right),\\
&\left|1\right>\aspace&\aspace=\mathbf{e}_1=\frac{\mathbf{e}_2\times\mathbf{e}_\mathbf{Q}}{|\mathbf{e}_2\times\mathbf{e}_\mathbf{Q}|}=
\frac1{\sqrt{Q_x^2+Q_y^2}\sqrt{Q_x^2+Q_y^2+Q_z^2}}\left(\begin{array}{c} -Q_xQ_z\\Q_yQ_z \\-Q_y^2-Q_x^2\end{array}\right),
\ea
and also the corresponding bra-vectors
\ba
&\left<1\right|\aspace&\aspace=
\frac{\left(-Q_xQ_z,Q_yQ_z,-Q_y^2-Q_x^2\right)}{\sqrt{Q_x^2+Q_y^2}\sqrt{Q_x^2+Q_y^2+Q_z^2}},\\
&\left<2\right|\aspace&\aspace=
\frac{\left(-Q_y, Q_x, 0\right)}{\sqrt{Q_x^2+Q_y^2}}.
\ea
For the degenerated case of $Q_x=0=Q_y$ we can regularize by choosing $Q_x=\iota$ and $Q_y=0.$ Then the limit $\iota\rightarrow 0$ gives the regularizations $\left|1\right>=\mathbf{e}_x$ and $\left|2\right>=\mathbf{e}_y.$
For all matrices $\mathsf{M}_{\alpha,\beta}$ where $\alpha,\,\beta= b,\,v$ we calculate the matrix elements in the two-dimensional space
\be
\left(\overline{\mathsf{M}}_{\alpha,\beta}\right)_{\mathrm{j}\,\mathrm{j}^\prime}=\left<\mathrm{j}\right|\mathsf{M}_{\alpha,\beta}\left|\mathrm{j}^\prime\right>,
\qquad \mbox{where}\;\;\mathrm{j},\,\mathrm{j}^\prime=1,\,2.
\ee
In such a way we obtain a reduced 4$\times$4 matrix
\be
\overline{\mathsf{M}} =
\left(\begin{array}{c|c}
 \overline{\mathsf{M}}_{bb} & \overline{\mathsf{M}}_{bv}\\
\hline
\overline{\mathsf{M}}_{vb} & \overline{\mathsf{M}}_{vv} \\
\end{array}\right)
\ee
whose eigenvectors are automatically perpendicular to $\mathbf{Q},$ simply because we have used the orthogonal to $\mathbf{Q}$ space.

As a rule the linearized analysis is made in Lagrangian, moving, wave-vector space
\be
\label{Psi_Lagrange}
\mathrm{d}_\tau\mathbf{K}(\tau)=\mathbf{U}_\mathrm{shear}(\mathbf{K}(\tau)),
\ee
with a time-dependent wave-vector
\be
K_x=K_{x,0}-K_y(\tau-\tau_0),\qquad K_y=\mathrm{const},\qquad K_z=\mathrm{const}
\ee
for each MHD wave.

In these coordinates for linearized waves the substantial time derivative $\mathrm{D}_\tau^{\mathrm{shear}}=\mathrm{d}_\tau$
is reduced to a usual time derivative and the separation of variables gives a set of ordinary independent equations for every MHD wave
\be
\mathrm{d}_\tau\Psi_\mathbf{K}(\tau)=\mathsf{M}(\mathbf{K}(\tau))\Psi_\mathbf{K}(\tau),
\qquad \mathbf{K}(\tau)\cdot\mathbf{v}_\mathbf{K}(\tau)=0,
\qquad \mathbf{K}(\tau)\cdot\mathbf{b}_\mathbf{K}(\tau)=0.
\ee
In this linearized case it is possible to exclude $b_z$ and $v_z$. In such a way we arrive at a simple-for-programming set of 4 equations
and 2 zero divergence conditions
\be
\mathrm{d}_\tau\psi_\mathbf{K}(\tau)=\mathsf{\tilde{M}}(\mathbf{K}(\tau))\psi_\mathbf{K}(\tau),
\qquad b_z=-(K_x(\tau)b_x+K_yb_y)/K_z,
\qquad v_z=-(K_x(\tau)v_x+K_yv_y)/K_z.
\ee

For small $K_y$ one can apply the WKB approximation supposing exponential time dependence of the MHD variables $\Psi(\tau)\propto \exp(\lambda\tau)$ and the wave amplitudes.
In the WKB approximation the energy amplification between $\tau=-\infty$ and $\tau=+\infty$ is given by the eigenvalue $\lambda$ with the maximal real part
\be
G\approx\exp\left(2\int_{-\infty}^{\infty}\mathrm{d}\tau\, \mathrm{Re}\,\lambda_{\mathrm{max}}(\mathbf{K}(\tau))\right).
\ee
For the case of MRI with nonzero $B_z$ the amplification factors are so giant that the linear analysis
makes no sense because the nonlinear terms become rather important and we have a nonlinear saturation of the MRI.
This saturation simulates strong turbulence for small wave-vectors, but definitely
for large wave-vectors $|K_y|\gg 1$ at $\tau \rightarrow\infty$ we have a wave type turbulence with a given frequency.

We have to mention that the linearized case of pure shear is exactly integrable in terms of the Heun functions~\cite{Mishonov:09,Mishonov:07}.
Investigating numerically this case with $\omega_{_\mathrm{C}}=0$ and $B_z=0$ in his Ph.D.\ thesis~\cite{Chagelishvili:93}
T.~Hristov discovered in 1990 the amplification of slow magnetosonic waves (SMWs) by shear flows.
Applied to the physics of accretion disks this amplification works even for purely azimuthal magnetic fields and gives a scenario for weak magnetic turbulence related to amplification of SMWs. We had to wait 30 years of incubation period, cf.~\cite{Dessler:70}, for the SMWs amplification to be recognized as an important issue for that astrophysical phenomenon.
In the next section we will consider how to proceed with the solution to the MHD equations.

\section{Energy density and power density} %
%
Our first step is to calculate the energy of plane MHD waves with time-dependent amplitudes.
Using that
\begin{equation}
 \int \mathrm{e}^{\ii\mathbf{Q}\cdot\mathbf{X}}\mathrm{d}^3X=(2\pi)^3\delta(\mathbf{Q})
\end{equation}
for the energy we obtain
\begin{equation}
\frac12\int\left(\rho\mathbf{V}^2_\mathrm{wave} + \frac1{\mu_0}\mathbf{B}^2_\mathrm{wave}\right)\mathrm{d}^3X=\rho V_\mathrm{A}^2
\sum_\mathbf{Q}\epsilon_{_\mathbf{Q}}, \qquad \epsilon_{_\mathbf{Q}}\equiv\frac12(\mathbf{v}_\mathbf{Q}^2+\mathbf{b}_\mathbf{Q}^2),
\end{equation}
i.e., the energy density is
\be
\rho V_\mathrm{A}^2 \int\int\int
\frac12\left[\mathbf{v}_\mathbf{Q}^2(\tau)+\mathbf{b}_\mathbf{Q}^2(\tau)\right]\,
\frac{\mathrm{d}Q_x\mathrm{d}Q_y\mathrm{d}Q_z}{(2\pi)^3}.
\ee
Analogously, with the help of the viscous stress tensor $\sigma^\prime_{ik}$ we express
the volume density of the wave heating
\begin{eqnarray}
Q^{\mathrm{wave}}_{\mathrm{kin}}\aspace&=&\aspace\int\sigma^\prime_{ik}\partial_k V^{\mathrm{wave}}_i \mathrm{d^3}x=
\frac12\int\sigma^\prime_{ik}(\partial_k V^{\mathrm{wave}}_i+\partial_iV^{\mathrm{wave}}_k)\mathrm{d^3}x=
\frac{\eta}{2}\int(\partial_k V^{\mathrm{wave}}_i+\partial_i V^{\mathrm{wave}}_k)^2\mathrm{d^3}x
\\\aspace&=&\aspace
\frac{\eta V_\mathrm{A}^2}{2\Lambda^2}\int\left(\sum_\mathbf{Q} Q_iv_k\mathrm{e}^{\ii\mathbf{Q}\cdot \mathbf{X}} + \sum_\mathbf{Q'} Q'_kv_i\mathrm{e}^{\ii\mathbf{Q'}\cdot \mathbf{X}} \right)^{\!\!2} \mathrm{d^3}x=
\rho AV^2_\mathrm{A}\nu'_\mathrm{k}\sum_Q Q^2 v_\mathbf{Q}^2. \nonumber
\end{eqnarray}

Similarly for the Ohmic part of the energy dissipation rate we have
\begin{equation}
Q^{\mathrm{wave}}_\mathrm{\Omega}=\mathbf{j}\cdot\mathbf{E}
=\frac{1}{\mu_0^2\sigma_{_{\Omega}}}(\mathrm{rot}\mathbf{B}_\mathrm{wave})^2
=\frac{B_0^2}{\mu_0^2\sigma_{_{\Omega}}}\left(\sum_\mathbf{Q} \nabla\times \mathbf{b}_\mathbf{Q}\mathrm{e}^{\ii\mathbf{Q}\cdot\mathbf{X}}\right)^{\!\!2}
=\rho AV^2_\mathrm{A}\nu'_\mathrm{m}\sum_\mathbf{Q} Q^2b^2_\mathbf{Q}.
\end{equation}
The dissipation rate of a laminar shear flow is given according to Newton's formula
\begin{equation}
Q^{\mathrm{shear}}_{\mathrm{kin}}=\frac{\eta}{2}\int(\partial_k V^{\mathrm{shear}}_i
+\partial_i V^{\mathrm{shear}}_k)^2\mathrm{d^3}x=\frac{\eta}{2}A^2(\delta_{k,x}\delta_{i,y}+\delta_{i,x}\delta_{k,y})^2=\eta A^2.
\end{equation}
Now we can calculate the total energy dissipation
$Q_\mathrm{tot}=Q^{\mathrm{shear}}_{\mathrm{kin}}+Q^{\mathrm{wave}}_{\mathrm{kin}}+Q^{\mathrm{wave}}_\mathrm{\Omega}$,
the viscosity and the effective viscosity
$\eta_\mathrm{eff}$
\begin{eqnarray}
\eta=\frac{Q^{\mathrm{shear}}_{\mathrm{kin}}}{A^2}, \qquad \eta_\mathrm{eff}=\rho\nu_\mathrm{eff}=\frac{Q_\mathrm{tot}}{A^2}.
\end{eqnarray}
In this way we can express the effective kinematic viscosity by the dimensionless Fourier components of the velocity and the magnetic field
\begin{eqnarray}
\nu_\mathrm{eff}(\tau)=\nu_\mathrm{k} + \nu_\mathrm{k}\sum_{\mathbf{Q}}Q^2\mathbf{v}_\mathbf{Q}^2(\tau)
+ \nu_\mathrm{m}\sum_{\mathbf{Q}} Q^2\mathbf{b}_\mathbf{Q}^2(\tau).
\end{eqnarray}
For example, if we have static probability distribution functions for the velocity and the magnetic field,
the enhancement factor of the effective viscosity is given by the time-averaged squares of the Fourier components for $\tau\gg1$
\begin{eqnarray}
\label{viscosity}
\frac{\eta_\mathrm{eff}}{\eta}=
1 + \sum_{\mathbf{Q}} Q^2\left<\mathbf{v}_\mathbf{ Q}^2\right>
+ \frac{\nu_\mathrm{m}}{\nu_\mathrm{k}}\sum_{\mathbf{ Q}} Q^2 \left<\mathbf{b}_\mathbf{Q}^2\right>;
\end{eqnarray}
this important parameter determines the work of the accretion discs as a machine for making of compact astrophysical objects.
The most simple scenario is to have the solution to the static equations for the $i$-th iteration of $\Psi_\mathbf{Q}$ and to calculate the next $(i+1)$-th iteration
\begin{eqnarray}
\partial_{\overline{\tau}}\mathbf{v}_{\mathbf{Q}}^{(i+1)}\aspace&=&\aspace
-Q_y\frac{\partial \mathbf{v}^{(\mathrm{i}+1)}_\mathbf{Q}}{\partial Q_x} =
-v^{(\mathrm{i}+1)}_{x,\mathbf{Q}}\mathbf{e}_y+2\frac{Q_y v^{(\mathrm{i}+1)}_{x,\mathbf{Q}}}{Q^2}\mathbf{Q}
+ 2\omega\left[\mathbf{n}(n_y v^{(\mathrm{i}+1)}_{x,\mathbf{Q}}-n_x
v^{(\mathrm{i}+1)}_{y,\mathbf{Q}})+(v^{(\mathrm{i}+1)}_{y,\mathbf{Q}}\mathbf{e}_x-v^{(\mathrm{i}+1)}_{x,\mathbf{Q}}\mathbf{e}_y)\right]  \nonumber\\
&&\aspace+(\bm{\alpha}\cdot\mathbf{Q})\mathbf{b}^{(\mathrm{i}+1)}_\mathbf{Q} -\nu^\prime_kQ^2\mathbf{v}^{(\mathrm{i}+1)}_\mathbf{Q}
+\Pi^{\perp\mathbf{Q}}\sum_{\mathbf{Q}'}\left[\mathbf{v}^{(\mathrm{i})}_{\mathbf{Q}'}\otimes\mathbf{v}^{(\mathrm{i})}_{\mathbf{Q}
-\mathbf{Q}'}\cdot\mathbf{Q'} +(\mathbf{Q}'\times\mathbf{b}^{(\mathrm{i})}_{\mathbf{Q}'})\times\mathbf{b}^{(\mathrm{i})}_{\mathbf{Q}-\mathbf{Q}'}\right],
\\
\partial_{\overline{\tau}}\mathbf{v}_{\mathbf{Q}}^{(i+1)}\aspace&=&\aspace-Q_y\frac{\partial \mathbf{b}^{(\mathrm{i}+1)}_\mathbf{Q}}{\partial Q_x} = b^{(\mathrm{i}+1)}_{x,\mathbf{Q}}\mathbf{e}_y
-(\mathbf{Q}\cdot\bm{\alpha})\mathbf{v}^{(\mathrm{i}+1)}_\mathbf{Q} -\nu'_\mathrm{m}Q^2\mathbf{b}^{(\mathrm{i}+1)}_\mathbf{Q}
-\mathbf{Q}\times\sum_{\mathbf{Q}'}(\mathbf{v}^{(\mathrm{i})}_{\mathbf{Q}'} \times \mathbf{b}^{(\mathrm{i})}_{\mathbf{Q}-\mathbf{Q}'}),\\
\partial_{\overline{\tau}}\aspace&\equiv&\aspace -Q_y\frac{\partial }{\partial Q_x}, \; \overline{\tau}=-\frac{Q_x}{Q_y},
\; \mathrm{D}_\tau=\partial_{\tau}+\partial_{\overline{\tau}},\; \mbox{for the independent variables ($\overline{\tau},Q_y,Q_z$}),\; Q_x=-Q_y\overline{\tau}.
\end{eqnarray}
For cold protoplanetary disks the Ohmic resistivity of weakly ionized gas is very high and the effective viscosity
is dominated in \Eqref{viscosity} by the $\nu_\mathrm{m}/\nu_\mathrm{k}$ term, in other words, the viscosity of the protoplanetary disks is created by Ohmic dissipation. Completely opposite is the situation for the hot almost completely-ionized hydrogen plasma in quasars.
The Ohmic resistivity is negligible and the effective viscosity is created by the Fourier components of the MHD waves $\left<\mathbf{v}_\mathbf{ Q}^2\right>.$
Only for small wave-vectors the MHD turbulence remains strong turbulence. At large wave-vectors
we have weak wave turbulence with wave-vectors going to infinity.
In the next section we will consider the stability conditions which have to be checked.

\section{Stability} %
%
The linear Lyapunov analysis which we outlined in Sec.~\ref{sec:Lyapunov} gives the idea what we have to do when we obtain
the static solution $\Psi_\mathbf{Q}^{(0)}=(\mathbf{b}_\mathbf{Q}^{(0)},\mathbf{v}_\mathbf{Q}^{(0)})$.
In order to investigate the stability of this static solution we have to consider a small time-dependent deviation from this solution
$\Psi_\mathbf{Q}^{(1)}(\tau)=(\mathbf{b}_\mathbf{Q}^{(1)}(\tau),\mathbf{v}_\mathbf{Q}^{(1)}(\tau))$.
In this case, neglecting the quadratic terms with respect to $\Psi_\mathbf{Q}^{(1)}$,
we find that the nonlinear terms in the MHD equations are linear integral operators in the $\mathbf{Q}$-space
\begin{equation}
\mathsf{\hat{N}'}\Psi_\mathbf{Q}^{(1)}=\left(\begin{array}{c}
\sum_{\mathbf{Q}'}\left[
\mathbf{Q}\cdot\mathbf{v}^{(1)}_{\mathbf{Q}'}\otimes\mathbf{v}^{(0)}_{\mathbf{Q}-\mathbf{Q}'}
+
\mathbf{Q}\cdot\mathbf{v}^{(0)}_{\mathbf{Q}'}\otimes\mathbf{v}^{(1)}_{\mathbf{Q}-\mathbf{Q}'}
+
(\mathbf{Q}'\times\mathbf{b}^{(0)}_{\mathbf{Q}'})\times\mathbf{b}^{(1)}_{\mathbf{Q}-\mathbf{Q}'}
+
(\mathbf{Q}'\times\mathbf{b}^{(1)}_{\mathbf{Q}'})\times\mathbf{b}^{(0)}_{\mathbf{Q}-\mathbf{Q}'}
\right]
\\
-\mathbf{Q}\times\sum_{\mathbf{Q}'}(
\mathbf{v}^{(0)}_{\mathbf{Q}'} \times \mathbf{b}^{(1)}_{\mathbf{Q}-\mathbf{Q}'}
+
\mathbf{v}^{(1)}_{\mathbf{Q}'} \times \mathbf{b}^{(0)}_{\mathbf{Q}-\mathbf{Q}'}
)
                 \end{array}\right).
\end{equation}
We obtain new terms in the eigenvalue problem which finally is reduced to the problem of obtaining the
maximal eigenvalue of an integral equation in which the coefficients are solutions to the static MHD equations.
Now let us analyze the perspectives.

\section{Perspective} %
%
The open question of the missing viscosity in accretion disks is a longstanding problem in physics.
In this work we have arrived at a complete set of numerically solvable ordinary differential equations which would help in finding a reasonable explanation of the aforementioned problem.

\begin{theacknowledgments}
The authors thank Dr.~Grigol Gogoberidze for his interest in the work and Prof.~Ivan Zhelyazkov for the critical
reading of the manuscript. This work was partially supported by the St.~Clement of Ohrid University at Sofia
Scientific Research Fund under Grant 203/2010.
\end{theacknowledgments}

\bibliographystyle{aipproc}   

\end{document}